\RequirePackage[l2tabu, orthodox]{nag}
\RequirePackage{snapshot}

\documentclass[10pt,twocolumn]{extarticle}

\makeatletter
\if@twocolumn
  \usepackage[dvips,letterpaper,top=0.5in, bottom=0.5in, left=0.75in, right=0.5in,includefoot,heightrounded]{geometry}
\else
  \usepackage[dvips,letterpaper,margin=1in,includefoot,heightrounded]{geometry}
\fi

\usepackage{srcltx}

\usepackage{amsmath}
\usepackage{amssymb,amsfonts}

\usepackage{abstract}

\usepackage{graphicx}
\usepackage{epsfig}
\usepackage[usenames,dvipsnames]{color}
\usepackage[usenames,dvipsnames]{xcolor}
\usepackage{subfigure}

\usepackage{booktabs}

\usepackage{setspace}
\usepackage{flushend}
\usepackage{multicol}

\usepackage{cite}
\usepackage{url}
\usepackage[normalem]{ulem}

\usepackage{enumerate}

\usepackage{multirow}

\usepackage{hyperref}

\usepackage[sc,tiny,center]{titlesec}

\usepackage{microtype}

\newtheorem{proposition}{Proposition}

\newtheorem{lemma}{Lemma}

\def\QED{\mbox{$\square$}}
\def\proof{\noindent{\it Proof:~}}
\def\endproof{\hspace*{\fill}~\QED\par\endtrivlist\unskip}

\newcommand{\dft}{\stackrel{\textsc{dft}}{\longleftrightarrow}}
\newcommand{\impulse}{\pmb{\delta}}
\newcommand{\Odd}{\operatorname{\mathcal{O}}}
\newcommand{\Even}{\operatorname{\mathcal{E}}}

\makeatletter

\newcommand{\printtitle}{%
\makeatletter
\if@twocolumn

\twocolumn[%
  \maketitle
  \begin{onecolabstract}
    \myabstract
  \end{onecolabstract}
  \begin{center}
    \small
    \textbf{Keywords}
    \\\medskip
    \mykeywords
  \end{center}
  \bigskip
]
\saythanks
\else
  \maketitle
  \begin{onecolabstract}
    \myabstract
  \end{onecolabstract}
  \begin{center}
    \small
    \textbf{Keywords}
    \\\medskip
    \mykeywords
  \end{center}
  \bigskip
  \onehalfspacing
\fi
\makeatother
}

\title{%
Multiuser Communication Based on the DFT Eigenstructure
}

\author{%
R.~M.~Campello de Souza
\thanks{%
R.~M.~Campello de Souza
is with
the
Departamento de Eletr\^onica e Sistemas,
Universidade Federal de Pernambuco.
E-mail:
\protect\url{ricardo@ufpe.br}
}
\and
H.~M.~de~Oliveira%
\thanks{%
H.~M.~de~Oliveira and R.~J.~Cintra
are with
the Signal Processing Group,
Departamento de Estat\'{\i}stica,
Universidade Federal de Pernambuco.
E-mail:
\protect\url{hmo@de.ufpe.br}, \protect\url{rjdsc@de.ufpe.br}
}
\and
R.~J.~Cintra${}^\dagger$%
}

\date{}

\newcommand{\myabstract}{%
The eigenstructure of the discrete Fourier transform (DFT) is examined
and
new systematic procedures to generate eigenvectors of the unitary DFT are proposed.
DFT eigenvectors are suggested as user signatures for data communication over the real adder channel (RAC).
The proposed multiuser communication system over the 2-user RAC is detailed.
}

\newcommand{\mykeywords}{%
DFT,
Eigenvectors,
Real adder channels
}

\begin{document}

\printtitle

\section{Introduction}

In general terms,
the eigenvectors
of a linear transformation~$\mathbf{T}$ are the vectors $\mathbf{v}$
that satisfy
$\mathbf{T}\,\mathbf{v} =\lambda \, \mathbf{v}$,
where $\lambda$ is a number called an eigenvalue.
The determination of eigenvectors and eigenvalues
constitute an appealing problem in many contexts~\cite{ref2,ref9,pei2004eigenvectors,pei2008generalized,ref3}. %

The unitary form of the discrete Fourier transform (DFT)~\cite{ref11} relates
an $N$-point complex vector
$\mathbf{x} = \big[x[n]\big]_{n=0}^{N-1}$
to a possibly complex vector
$\mathbf{X} = \big[X[k]\big]_{k=0}^{N-1}$
according to
$\mathbf{X}=\frac{1}{\sqrt{N}}\,\mathbf{W}\,\mathbf{x}$,
where the transformation matrix is specified by
$\mathbf{W} = [\omega^{-nk}]_{n,k=0}^{N-1}$
and
$\omega = \exp(j2\pi/N)$ is
an
$N$th root of unity.
This formulation of the DFT avoids cumbersome asymmetrical constants
and
is algebraically equivalent to other instantiations.
The resulting DFT pair is denoted by $\mathbf{x}\dft \mathbf{X}$.

The eigenstructure of the DFT was early investigated in~\cite{ref3,ref4}.
From a theoretical point of view,
various linear transformations,
such as the generalized DFT and other discrete transforms,
have also had their eigenstructure examined~\cite{ref9,ref10}. %

Closed forms for particular classes of offset DFT eigenvectors
were analyzed in~\cite{pei2004eigenvectors},
furnishing theoretical connections to the eigenvectors of
the discrete cosine transform and the discrete Hartley transform.
Moreover,
several methods for the generation of DFT eigenvectors
are based on commuting matrices. %
In recent works~\cite{pei2009arbitrary,pei2008generalized},
comprehensive studies of commuting matrices are described.
In a comparable framework,
eigenvectors of discrete transforms have been studied and applied
in (i)~chirp filtering methods~\cite{ozaydin2006orthogonal};
(ii)~watermarking techniques~\cite{pei2004eigenvectors};
and
(iii)~motion blur parameter identification~\cite{yoshida1993blur},
to cite but a few areas.

The aim of this paper is twofold.
Firstly,
the DFT eigenvectors
are examined.
Such vectors
are termed invariant sequences.
Since an invariant sequence~$\mathbf{x}$
satisfies $\mathbf{X}=\lambda\mathbf{x}$,
its DFT spectrum can be trivially obtained.
Secondly,
new communication systems
based on the DFT eigenstructure
are introduced,
being the main goal of the paper.
The mathematical framework of the eigenvectors and eigenvalues
of the DFT furnishes the necessary tools
for the design of multiuser systems over
real adder channels (RAC).
In particular,
communication schemes for the 2-, 3-, and 4-user RAC are proposed.

Henceforth,
the term sequence is used interchangeably with vector.
All sequences considered are periodic with period~$N$
and index manipulation follows a modulo-$N$ arithmetic.
Additionally,
sequences are represented by column vectors
and
multiplications between vectors are entrywise operations.

The paper is organized as follows.
Section~\ref{section.eigensequences.dft}
introduces new systematic procedures to generate DFT eigenvectors.
A new multiuser communication system, for 2-, 3-, and 4-users,
is proposed in Section~\ref{section.sequences.for}.
In Section~\ref{section.design}, the basic ideas for the design of a 2-user
communication system are elaborated.
Multiuser systems are also addressed.
Section~\ref{section.conclusions} summarizes the results obtained.

\section{A Generator of Invariant Sequences}
\label{section.eigensequences.dft}

The DFT structure severely restrains the possible values of $\lambda$
as shown in the following results due to McClellan-Parks~\cite{ref3}.

\begin{lemma}%
\label{prop1}
The eigenvalues of the unitary DFT are the fourth roots of unity: $\pm1$, $\pm j$ .
\end{lemma}

\begin{lemma}%
\label{prop2}
According as $\lambda = \pm1$ or $\lambda=\pm j$,
the invariant sequences possess even or odd symmetry, respectively.
\end{lemma}

In the remaining of this section, two methods for the generation of invariant sequences are suggested.
The first procedure explores the circular nature of the DFT
and
the second one utilizes the convolution theorem.

Using the fact that
the fourth successive application of the DFT
to
a given sequence
returns the original sequence itself,
the following proposition can be obtained.

\begin{proposition}%
\label{prop.method.1}
Let $\mathbf{x}$ be a given sequence.
Then $\mathbf{x}+\lambda\mathbf{X}+\lambda^2\mathbf{x}^-+\lambda^3\mathbf{X}^-$
is an invariant sequence,
for $\lambda\in\{ \pm1, \pm j\}$.
\end{proposition}
\proof
Noticing that
\begin{align*}
\mathbf{x}
\dft
\mathbf{X}
\dft
\mathbf{x}^-
\dft
\mathbf{X}^-
\dft
\mathbf{x},
\end{align*}
where
$\mathbf{x}^- = \big[x[N-n]\big]_{n=0}^{N-1}$
and
$\mathbf{X}^- = \big[X[N-k]\big]_{k=0}^{N-1}$,
and applying the linearity property of the DFT,
the result follows.
\endproof

The convolution theorem offers a more general approach to invariant sequences construction
as shown below.

\begin{proposition}
\label{prop.general}
Let $\mathbf{x}$ and $\mathbf{y}$ be arbitrary sequences.
For $\lambda\in\{ \pm1, \pm j\}$,
the sequence given by
\begin{align*}
\mathbf{x}\ast\mathbf{y}
+
\lambda\sqrt{N}\mathbf{X}\mathbf{Y}
+
\lambda^2\mathbf{x}^-\ast\mathbf{y}^-
+
\lambda^3\sqrt{N}\mathbf{X}^-\mathbf{Y}^-,
\end{align*}
is an invariant sequence,
where
$\mathbf{x}\dft\mathbf{X}$,
$\mathbf{y}\dft\mathbf{Y}$,
and $\ast$ denotes the cyclic convolution operation.
\end{proposition}
\proof
The convolution theorem asserts that
$\mathbf{x}\ast\mathbf{y} \dft \sqrt{N} \mathbf{X}\mathbf{Y}$.
Thus,
considering the duality property of the DFT,
it follows that
\begin{gather*}
\mathbf{x}\ast\mathbf{y}
+
\lambda\sqrt{N}\mathbf{X}\mathbf{Y}
+
\lambda^2\mathbf{x}^-\ast\mathbf{y}^-
+
\lambda^3\sqrt{N}\mathbf{X}^-\mathbf{Y}^-
\\
\dft
\\
\sqrt{N}\mathbf{X}\mathbf{Y}
+
\lambda
\mathbf{x}^-\ast\mathbf{y}^-
+
\lambda^2
\sqrt{N}\mathbf{X}^-\mathbf{Y}^-
+
\lambda^3\mathbf{x}\ast\mathbf{y}.
\end{gather*}
Since $\lambda^4=1$, the transformed sequence is equal to
\begin{align*}
\lambda^3
\left(
\mathbf{x}\ast\mathbf{y}
+
\lambda\sqrt{N}\mathbf{X}\mathbf{Y}
+
\lambda^2\mathbf{x}^-\ast\mathbf{y}^-
+
\lambda^3\sqrt{N}\mathbf{X}^-\mathbf{Y}^-
\right),
\end{align*}
which concludes the proof.
\endproof

Proposition~\ref{prop.general} can also be used to construct families
of invariant sequences associated with a particular invariant sequence.
Indeed, generating functions for invariant sequences can be obtained
according to the following result.

\begin{proposition}%
\label{prop.generating}
Let $\mathbf{x}$ be an invariant sequence.
Given an arbitrary sequence $\mathbf{y}$,
for $\lambda\in\{\pm1, \pm j\}$,
families of invariant sequences associated with $\mathbf{x}$ can
be derived by the generating function
\begin{align*}
\mathbf{g}
=
\mathbf{x}\ast\mathbf{y}
+
\lambda\lambda'
\sqrt{N}
\mathbf{x}\mathbf{Y}
+
\lambda^2\mathbf{x}^-\ast\mathbf{y}^-
+
\lambda^3\lambda'
\sqrt{N}
\mathbf{x}^-\mathbf{Y}^-,
\end{align*}
where $\lambda'$ is the eigenvalue associated with $\mathbf{x}$.
\end{proposition}
\proof
It follows from Proposition~\ref{prop.general} for $\mathbf{X}=\lambda'\mathbf{x}$.
\endproof

Let the delayed unit sample sequence $\impulse_m$ be
a sequence of null components except for the $m$th one which is unitary.
By letting $\mathbf{y}=\impulse_m$,
a particular class of generating functions is derived as follows:
\begin{align*}
\mathbf{g}_m
=
\mathbf{x}\ast\impulse_m
+
\lambda\lambda'
\mathbf{x}\mathbf{w}_{-m}
+
\lambda^2
\mathbf{x}^-\ast\impulse_{-m}
+
\lambda^3\lambda'
\mathbf{x}^- \mathbf{w}_{m},
\end{align*}
where
$m=0,1,\ldots,N-1$,
and
$\mathbf{w}_m = [ \omega^{mn} ]_{n=0}^{N-1}$
is the exponential sequence,
which is the DFT of $\sqrt{N}\impulse_{m}$.
Despite the multiple combinations of values for
$\lambda$ and $\lambda'$,
only four distinct generating functions are obtained:
\begin{align*}
\mathbf{g}_m^{(0)}
=
\mathbf{x}\ast(\impulse_m + \impulse_{-m})
+
2\mathbf{x}\,\mathbf{cos}_m,  \\
\mathbf{g}_m^{(1)}
=
\mathbf{x}\ast(\impulse_m - \impulse_{-m})
+
2\mathbf{x}\,\mathbf{sin}_m, \\
\mathbf{g}_m^{(2)}
=
\mathbf{x}\ast(\impulse_m + \impulse_{-m})
-
2\mathbf{x}\,\mathbf{cos}_m,  \\
\mathbf{g}_m^{(3)}
=
\mathbf{x}\ast(\impulse_m - \impulse_{-m})
-
2\mathbf{x}\,\mathbf{sin}_m,
\end{align*}
where
the cosine and sine sequences are defined as
$\mathbf{cos}_m = \frac{\mathbf{w}_m+\mathbf{w}_{-m}}{2}$
and
$\mathbf{sin}_m = \frac{\mathbf{w}_m-\mathbf{w}_{-m}}{2j}$,
respectively.

More sophisticated expressions for the generating functions can be derived by letting
$\mathbf{y} = \sum_{n=0}^{N-1} y[n] \impulse_{n}$,
where $y[n], n=0,1,\ldots,N-1$ are the components of $\mathbf{y}$.
Due to the distributive properties of both convolution and multiplication operations,
it is possible to write
\begin{align*}
\mathbf{g}
=
\sum_{n=0}^{N-1}
y[n]
\mathbf{g}_n^{(k)},
\quad
k=0,1,2,3,
\end{align*}
the matrix form of which is
$\mathbf{g}=\mathbf{G}^{(k)}\mathbf{y}$,
where
$
\mathbf{G}^{(k)}
=
\begin{bmatrix}
\mathbf{g}_0^{(k)}
&|&
\mathbf{g}_2^{(k)}
&|&
&\cdots&
&|&
\mathbf{g}_{N-1}^{(k)}
\end{bmatrix}
$
is called the eigenvector mapping matrix.
Thus, any sequence $\mathbf{y}$ can be converted into an invariant sequence.

The space of finite energy sequences
can be split into four orthogonal
eigenspaces,
$V_{+1}$, $V_{-1}$, $V_{+j}$, and $V_{-j}$,
associated with the eigenvalues
$+1$, $-1$, $+j$, and $-j$,
respectively~\cite{dym1972}.
Moreover,
the eigenvector mapping matrices
can assign eigenvectors from one eigenspace to another.
The eigenspace of $\mathbf{g}$
depends on the eigenspace where $\mathbf{x}$ is
defined.
If $\mathbf{x}\in V_\lambda$,
for $\lambda \in \{\pm1, \pm j\}$,
then
the sequence
$\mathbf{g} = \mathbf{G}^{(k)}\mathbf{y} \in V_{\lambda j^k}$
for $k=0,1,2,3$.

For instance,
let $\mathbf{x} = \impulse_0 + \mathbf{w}_0/\sqrt{N}$, which is one of the simplest invariant sequences
with $\lambda'=1$.
For illustrative purposes,
Table~\ref{tab1}
lists,
for selected values of $N$,
some invariant sequences
generated by
$\mathbf{g}_m^{(0)}$,
$\mathbf{g}_m^{(1)}$,
$\mathbf{g}_m^{(2)}$,
and
$\mathbf{g}_m^{(3)}$.

\begin{table*}
\centering
\caption{Selected invariant sequences}
\label{tab1}
\begin{tabular}{crl}%
\hline
$N$ & $\lambda$ & \multicolumn{1}{c}{Invariant sequence} \\
\hline
\multirow{4}{*}{$6$}&
$1$&
$2+\frac{2\sqrt{6}}{3}$, $1+\frac{\sqrt{6}}{2}$, $\frac{\sqrt{6}}{6}$, $0$, $\frac{\sqrt{6}}{6}$,  $1+\frac{\sqrt{6}}{2}$
\\
& $-1$&
$-2$, $1+\frac{\sqrt{6}}{6}$, $\frac{\sqrt{6}}{2}$, $\frac{2\sqrt{6}}{3}$, $\frac{\sqrt{6}}{2}$, $1+\frac{\sqrt{6}}{6}$
\\
& $j$&
$0$, $1-\frac{\sqrt{2}}{2}$, $-\frac{\sqrt{2}}{2}$, $0$, $\frac{\sqrt{2}}{2}$, $-1+\frac{\sqrt{2}}{2}$
\\
& $-j$&
$0$, $1+\frac{\sqrt{2}}{2}$, $\frac{\sqrt{2}}{2}$, $0$, $-\frac{\sqrt{2}}{2}$, $-1-\frac{\sqrt{2}}{2}$
\\
\hline
\multirow{4}{*}{$8$}&
$1$&
$2+\sqrt{2}$,   $\frac{3+\sqrt{2}}{2}$,  $\sqrt{2}/2$,   $\frac{-1+\sqrt{2}}{2}$,    $0$,   $\frac{-1+\sqrt{2}}{2}$,   $\sqrt{2}/2$,  $\frac{3+\sqrt{2}}{2}$
\\
& $-1$&
$-2$,   $\frac{1+\sqrt{2}}{2}$,   $\sqrt{2}/2$,   $\frac{1+\sqrt{2}}{2}$,   $\sqrt{2}$,   $\frac{1+\sqrt{2}}{2}$,   $\sqrt{2}/2$, $\frac{1+\sqrt{2}}{2}$
\\
& $j$&
$0$,  $1/2$,  $-\sqrt{2}/2$,  $-1/2$, $0$,  $1/2$,  $\sqrt{2}/{2}$,  $-1/2$
\\
& $-j$&
$0$,  $3/2$,  $\sqrt{2}/2$,   $1/2$,   $0$,  $-1/2$,  $-\sqrt{2}/2$, $-3/2$
\\
\hline
\end{tabular}
\end{table*}

\section{Sequences for the Real Adder Channel}
\label{section.sequences.for}

A well-known communications channel model is the two-user binary adder channel, or 2-user BAC~\cite{ref13}.
In this paper,
the 2-, 3-, and 4-user RAC models are focused,
where addition is over the real numbers.
Invariant sequences
can be used
for information transmission of two to four users over the RAC.
In the following,
for $i=1,2,3,4$,
let $\mathbf{x}_i$ be an invariant sequence with DFT given by $\mathbf{X}_i$.

\subsection{The 2-user RAC}

Let $\mathbf{x}_1$ and $\mathbf{x}_2$ be
defined in $V_{+1}$  and $V_{-1}$
and
associated with users 1 and 2, respectively.
The 2-user RAC furnishes a new sequence $\mathbf{y} = \mathbf{x}_1+\mathbf{x}_2$
from which
individual user sequences can be discriminated.
Denoting the DFT pair
$\mathbf{y}\dft \mathbf{Y}$, we may write
\begin{align}
\mathbf{y}&=\mathbf{x}_1+\mathbf{x}_2, \label{eq4a} \\
\mathbf{Y}&=\mathbf{X}_1+\mathbf{X}_2. \label{eq4b}
\end{align}
Since $\mathbf{x}_1$ and $\mathbf{x}_2$ are in different eigenspaces,
$V_{+1}$ and $V_{-1}$, respectively,
(\ref{eq4b}) is equivalent to
\begin{align}
\label{eq5}
\mathbf{Y}=\mathbf{x}_1-\mathbf{x}_2.
\end{align}
From (\ref{eq4a}) and~(\ref{eq5}),
user sequences may be recovered according to
$\mathbf{x}_1=\frac{\mathbf{y}+\mathbf{Y}}{2}$
and
$\mathbf{x}_2=\frac{\mathbf{y}-\mathbf{Y}}{2}$,
where FFT procedures can be used to compute the DFT of $\mathbf{y}$.
Considering the four eigenvalues,
a total of six similar systems can be implemented.
For each one, expressions for $\mathbf{x}_1$ and $\mathbf{x}_2$ can be directly calculated
and are given in Table~\ref{tab3.2}.

\begin{table}
\centering
\caption{User recovery expressions for the 2-user RAC}
\begin{tabular}{ccc}
\hline
Eigenspace selection & User 1 & User 2 \\
\hline
$(V_{+1},V_{-1})$
&
$\frac{\mathbf{y}+\mathbf{Y}}{2}$
&
$\frac{\mathbf{y}-\mathbf{Y}}{2}$
\\
$(V_{+1},V_{+j})$
&
$\frac{-\mathbf{y}+\mathbf{Y}}{1-j}$
&
$\frac{\mathbf{y}-\mathbf{Y}}{1-j}$
\\
$(V_{+1},V_{-j})$
&
$\frac{j\mathbf{y}+\mathbf{Y}}{1+j}$
&
$\frac{j\mathbf{y}-\mathbf{Y}}{1+j}$
\\
$(V_{-1},V_{+j})$
&
$\frac{j\mathbf{y}-\mathbf{Y}}{1+j}$
&
$\frac{\mathbf{y}+\mathbf{Y}}{1+j}$
\\
$(V_{-1},V_{-j})$
&
$\frac{-j\mathbf{y}-\mathbf{Y}}{1-j}$
&
$\frac{\mathbf{y}+\mathbf{Y}}{1-j}$
\\
$(V_{+j},V_{-j})$
&
$\frac{\mathbf{y}-j\mathbf{Y}}{2}$
&
$\frac{\mathbf{y}+j\mathbf{Y}}{2}$
\\
\hline
\end{tabular}
\label{tab3.2}
\end{table}

\subsection{The 3-user RAC}

The three-user real adder channel provides the
output signal $\mathbf{y}=\mathbf{x}_1+\mathbf{x}_2+\mathbf{x}_3$.
Judiciously assigning to each user an invariant sequence extracted
from mutually different eigenspaces,
user sequences can be recovered from $\mathbf{y}$.
The four possible eigenspaces give four choices of design,
corresponding to the following sets of eigenvalues:
$\{+1, -1, +j\}$,
$\{+1, -1, -j\}$,
$\{+1, +j, -j\}$,
and
$\{-1, +j, -j\}$.

Let us turn our attention to the
design associated with the
$\{+1, -1, +j\}$ selection of eigenstructure.
By applying twice the DFT to $\mathbf{y}$,
the set of equations below arises:
\begin{align*}
\mathbf{x}_1 + \mathbf{x}_2 + \mathbf{x}_3 &= \mathbf{y}, \\
\mathbf{x}_1 - \mathbf{x}_2 + j\mathbf{x}_3 &=  \mathbf{Y},  \\
\mathbf{x}_1 + \mathbf{x}_2 - \mathbf{x}_3 &= \mathbf{y}^-,
\end{align*}
the solution of which is promptly expressed by
$\mathbf{x}_1 = \frac{\Even\{\mathbf{y}\} - j\Odd\{\mathbf{y}\} + \mathbf{Y}}{2}$,
$\mathbf{x}_2 = \frac{\Even\{\mathbf{y}\} + j\Odd\{\mathbf{y}\} - \mathbf{Y}}{2}$,
and
$\mathbf{x}_3 = \Odd\{\mathbf{y}\}$,
where
operators $\Even\{\cdot\}$ and $\Odd\{\cdot\}$
denote the even and odd part of their arguments, respectively.

The remaining cases are completely analogous and
the user recovery expressions for other eigenspace selections
are indicated in Table~\ref{tab3.3}.

\begin{table*}
\centering
\caption{User recovery expressions for the 3-user RAC}
\label{tab3.3}
\begin{tabular}{cc}
\hline
Eigenspace selection & User 1; User 2; User 3\\
\hline
$(V_{+1},V_{-1}, V_{+j})$
&
$\frac{\Even\{\mathbf{y}\} - j\Odd\{\mathbf{y}\} + \mathbf{Y}}{2}$
;
$\frac{\Even\{\mathbf{y}\} + j\Odd\{\mathbf{y}\} - \mathbf{Y}}{2}$
;
$\Odd\{\mathbf{y}\}$
\\
$(V_{+1},V_{-1}, V_{-j})$
&
$\frac{\Even\{\mathbf{y}\} + j\Odd\{\mathbf{y}\} + \mathbf{Y}}{2}$
;
$\frac{\Even\{\mathbf{y}\} - j\Odd\{\mathbf{y}\} - \mathbf{Y}}{2}$
;
$\Odd\{\mathbf{y}\}$
\\
$(V_{+1}, V_{+j}, V_{-j})$
&
$\Even\{\mathbf{y}\}$
;
$\frac{\Odd\{\mathbf{y}\} + j\Even\{\mathbf{y}\} -j \mathbf{Y}}{2}$
;
$\frac{\Odd\{\mathbf{y}\} - j\Even\{\mathbf{y}\} +j \mathbf{Y}}{2}$
\\
$(V_{-1}, V_{+j}, V_{-j})$
&
$\Even\{\mathbf{y}\}$
;
$\frac{\Odd\{\mathbf{y}\} - j\Even\{\mathbf{y}\} -j \mathbf{Y}}{2}$
;
$\frac{\Odd\{\mathbf{y}\} + j\Even\{\mathbf{y}\} +j \mathbf{Y}}{2}$
\\
\hline
\end{tabular}
\end{table*}

\subsection{The 4-user RAC}

The 4-user RAC admits
only one design option.
In fact, all eigenstructures must be considered simultaneously
and each user is assigned an eigenvector from a distinct eigenspace.
The output signal is given by
$\mathbf{y} = \mathbf{x}_1+\mathbf{x}_2+\mathbf{x}_3+\mathbf{x}_4$.
Users can be recovered from $\mathbf{y}$
according to similar manipulations employed for
the 2- and 3-user RAC cases.
Therefore,
after three successive applications of the DFT to
$\mathbf{y}$,
the following system of equations is obtained:
\begin{align*}
\mathbf{x}_1+\mathbf{x}_2+\mathbf{x}_3+\mathbf{x}_4 &= \mathbf{y}, \\
\mathbf{x}_1-\mathbf{x}_2+j\mathbf{x}_3-j\mathbf{x}_4 &= \mathbf{Y}, \\
\mathbf{x}_1+\mathbf{x}_2-\mathbf{x}_3-\mathbf{x}_4 &= \mathbf{y}^-, \\
\mathbf{x}_1-\mathbf{x}_2-j\mathbf{x}_3+j\mathbf{x}_4 &= \mathbf{Y}^-.
\end{align*}
The solution for the above system is expressed by
$\mathbf{x}_1 = \frac{\Even\{\mathbf{y} \} + \Even\{\mathbf{Y} \}}{2}$,
$\mathbf{x}_2 = \frac{\Even\{\mathbf{y} \} - \Even\{\mathbf{Y} \}}{2}$,
$\mathbf{x}_3 = \frac{\Odd\{\mathbf{y} \} -j \Odd\{\mathbf{Y} \}}{2}$,
and
$\mathbf{x}_4 = \frac{\Odd\{\mathbf{y} \} +j \Odd\{\mathbf{Y} \}}{2}$.

\section{Design of New Communication Systems}
\label{section.design}

\subsection{A Two-user System}

Without any loss of generality,
a 2-user transmission system over the RAC
is detailed.
As described below,
the communication systems for the 3- and 4-user RAC
follow
similar principles and
design issues.

Let the symbols $a_1$ and $a_2$ be real numbers that
convey the information generated by two users.
The generated data are linearly combined in the RAC,
resulting in
$\mathbf{y} = a_1 \mathbf{x}_1 + a_2 \mathbf{x}_2$,
where $\mathbf{x}_1$ and $\mathbf{x}_1$ and invariant sequences.

Any two invariant sequences may be assigned to the users,
as long as the sequences belong to different eigenspaces.
In particular,
a possible design is to ensure that
$\mathbf{x}_1\in V_{+1}$ and
$\mathbf{x}_2\in V_{-1}$.
For digital transmission of information,
the binary sequences provided by
each user are split into $b$-bit words.
Subsequently,
a transmission interval $T$ is selected
to match
the duration of each sequence $\mathbf{x}_i$,
and
the user transmission rate is $\text{$b$ bits}/ T$.
A digital-to-analog converter (D/A)
maps the user bit strings into coefficients $a_1$ and $a_2$
according to $2^b$ quantization levels of the considered dynamic range,
where $b$ is the D/A resolution in bits.

\begin{figure}
\centering
\subfigure[]{\epsfig{file=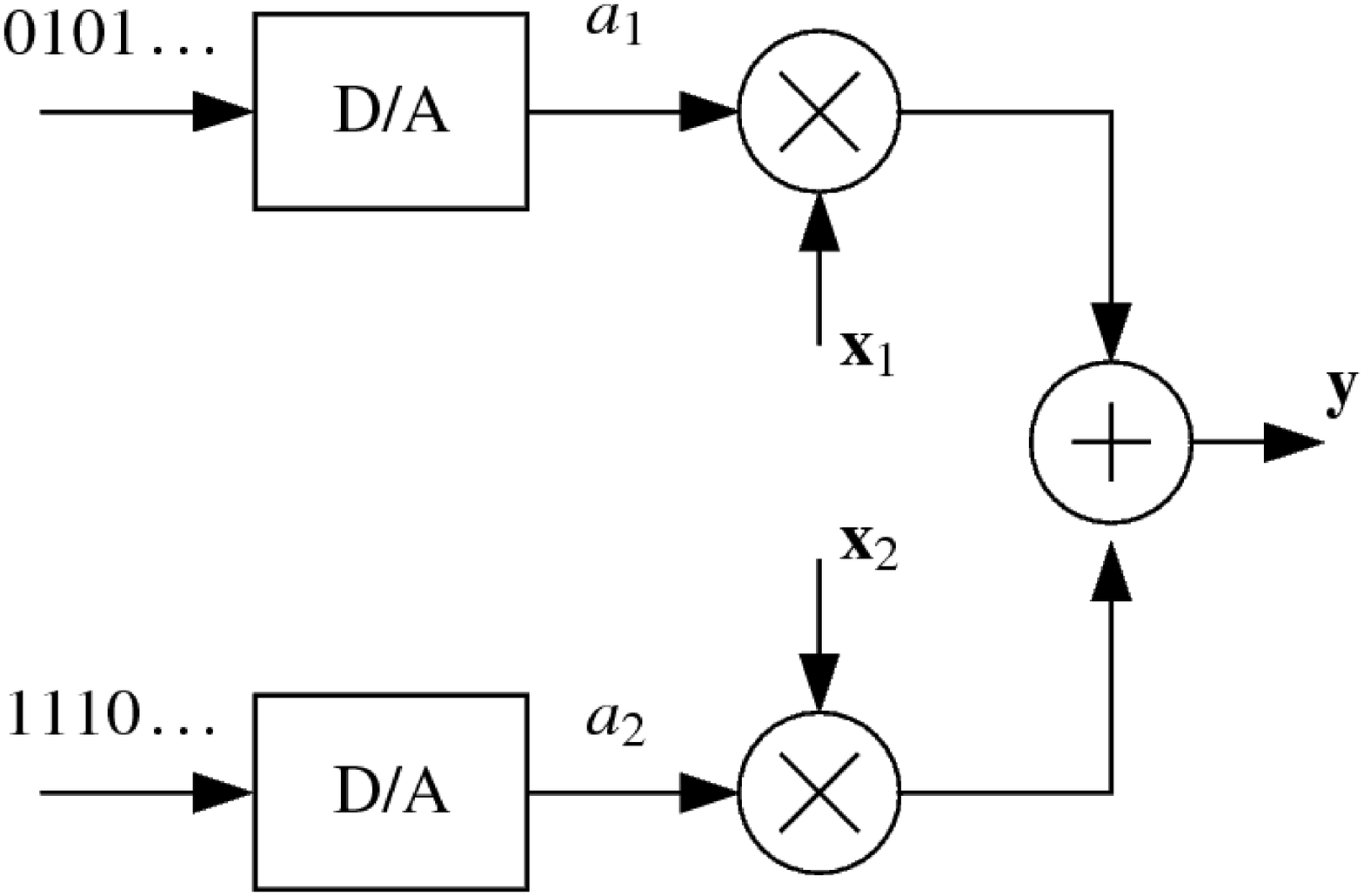,width=5cm}}
\subfigure[]{\epsfig{file=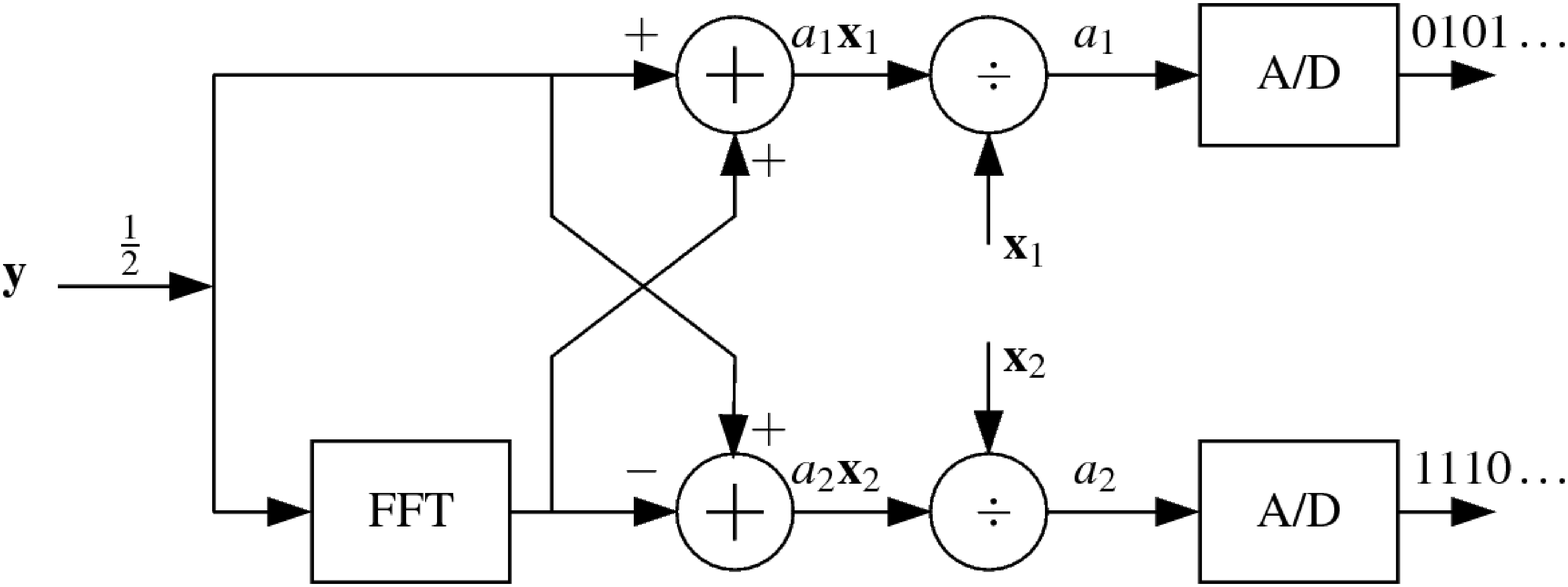,width=7.5cm}}
\caption{Possible transmitter (a) and receiver (b) diagrams for the 2-user RAC.}
\label{fig1}
\end{figure}

The receiver side performs the operations described in
Section~\ref{section.sequences.for}
as well as the analog-to-digital conversion~(A/D).
Figure~\ref{fig1} depicts
the block diagrams for both the transmitter and the receiver
parts of the discussed transmission system over the 2-user RAC.

Depending on signal-to-noise ratio conditions,
the computations at the receiver side can be
significantly reduced.
For example,
consider the scenario furnished of the 2-user system over the noiseless channel.
The values of
$a_1$ and $a_2$ can be obtained from
$a_1 = \frac{(y[0]+Y[0])/2}{x_1[0]}$
and
$a_2 = \frac{(y[0]-Y[0])/2}{x_2[0]}$.
Observe that $Y[0]$ is simply the mean value of $\mathbf{y}$
scaled by $1/\sqrt{N}$.
Thus,
the entire computation for $a_1$ and $a_2$
requires
$N+1$ additions (partial evaluation of $Y[0]$ and butterfly overhead),
two divisions by 2 (bitwise shift operations),
and
three multiplications ($1/x_1[0]$, $1/x_2[0]$, and $1/\sqrt{N}$ due to the computation of $Y[0]$).

\subsection{A Multiuser System}

A multiuser system can be devised.
Users are divided into 2, 3 or 4 groups
and
each group is assigned to a particular eigenspace
$V_{+1}$,
$V_{-1}$,
$V_{+j}$,
or
$V_{-j}$.
Users of a given group receive distinct invariant sequences,
which belongs to the eigenspace associated with the group.
The assigned sequences are interpreted as user signature sequences.

\begin{figure}
\centering
{\epsfig{file=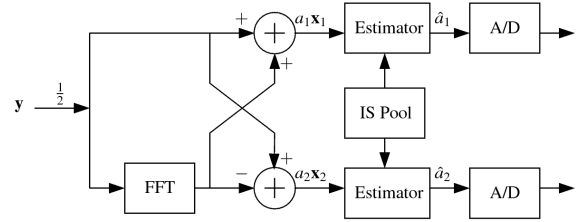,width=7.5cm}}
\caption{Possible receiver diagram for the discussed multiuser system.}
\label{fig2}
\end{figure}

Although the receiver has access to the set of all possible
invariant sequences applicable for a particular system design,
it has no \emph{a priori} information of
which invariant sequence was selected for transmission.
As a result,
the receiver must estimate the necessary invariant sequence
from the received data.
In this case,
Figure~\ref{fig2} illustrates the block diagram for
the 2-user system.

A well-known estimator is the sample mean.
It can be employed to obtain an estimate of the
transmitted data $a_k$, say $\hat{a}_k$,
according to the following algorithm:
\begin{enumerate}[(i)]
\item
For all $\mathbf{x}_i$ from an invariant sequence pool,
calculate the sequences $\mathbf{z}_i =\frac{a_k\mathbf{x}_k}{\mathbf{x}_i}$;

\item
Compute
$\hat{k}
=
\mathrm{arg} \min_i \mathrm{var}\{\mathbf{z}_i\}$,
where $\mathrm{var}\{\cdot\}$ is the variance operator;

\item
The estimated transmitted data is given by
$\hat{a}_k = \mathrm{mean}\{\mathbf{z}_{\hat{k}}\}$,
where $\mathrm{mean}\{\cdot\}$ returns the
sample mean of its argument.

\end{enumerate}

\subsection{Energy Considerations}

In a practical scenario,
the maximum allowed channel energy, $E_\text{max}$,
can determine the dynamic range of the user coefficients.
Assume that
$|a_i| \leq M$, $i=1,2$, where $M$ is a finite real quantity.
A sufficient condition for the choice of the peak value $M$ is established in the following proposition.

\begin{proposition}
\label{energy}
Consider a 2-user RAC,
where the sequences energy is limited by a quantity $E_\text{max}$.
If user sequences $\mathbf{x}_1 \in V_{+1}$ and $\mathbf{x}_2 \in V_{-1}$
are weighted by scalars $a_1$, $a_2$, respectively, with dynamic range
\begin{align*}
\pm
\sqrt
\frac{E_\text{max}}%
{\| \mathbf{x}_1 \|^2 + \|\mathbf{x}_2  \|^2},
\end{align*}
then the channel sequence energy does not exceed $E_\text{max}$.
The symbol $\|\cdot \|$ denotes the usual 2-norm of its argument.
\end{proposition}
\proof
Considering that
$|a_i|\leq M$,
an application of the triangle inequality
gives
an upper bound for the energy of the channel
sequence $\mathbf{y}=a_1\mathbf{x}_1+a_2\mathbf{x}_2$:
\begin{align*}
\|\mathbf{y}\|^2
& =
\| a_1\mathbf{x}_1+a_2\mathbf{x}_2  \|^2
\leq
\| a_1\mathbf{x}_1 \|^2 + \|a_2\mathbf{x}_2  \|^2 \\
&\leq
M^2 \left( \| \mathbf{x}_1 \|^2 + \|\mathbf{x}_2  \|^2 \right)
=
E_\text{max}.
\end{align*}
\endproof

In a noiseless channel,
$b$ may assume an arbitrarily large value,
and the transmission rate is unbounded.
The suggested transmission system can be interpreted
as a generalization of direct sequence spread spectrum systems,
e.g., DS-CDMA.
Instead of spreading one bit using a signature sequence,
$b$ bits are spread over a length $N$ sequence.

\section{Conclusions}
\label{section.conclusions}

This paper investigates sequences that are invariant to the unitary DFT.
Systematic procedures for the generation of these sequences were proposed.
In particular,
generating functions for invariant sequences of arbitrary length were introduced.
A multiuser communication system for the real adder channel,
for 2, 3 and 4 users was proposed.
The suggested system applies DFT invariant sequences as user signatures.
The proposed signatures are non-binary, in contrast to the binary spread
spectrum sequences.
The performance of the system in both the noiseless and noisy
cases are currently under investigation.

\section*{Acknowledgments}

This work was partially supported by
CNPq (Brazil)
and
DFAIT (Canada).

{\small
\bibliographystyle{abbrv}
\bibliography{IEEEabrv,ref}

\begin{thebibliography}{10}

\bibitem{ref13}
R.~Ahlswede and V.~B. Balakirsky.
\newblock Construction of uniquely decodable codes for the two-user binary
  adder channel.
\newblock {\em {IEEE} Trans. Inf. Theory}, 45(1):326--330, Jan. 1999.

\bibitem{ref10}
G.~Cariolaro, T.~Erseghe, and P.~Kraniauskas.
\newblock The fractional discrete cosine transform.
\newblock {\em {IEEE} Trans. Signal Process.}, 50:902--911, Apr. 2002.

\bibitem{dym1972}
H.~Dym and H.~P. McKean.
\newblock {\em Fourier Series and Integrals}.
\newblock Academic Press, 1972.

\bibitem{ref11}
D.~F. Elliott and K.~R. Rao.
\newblock {\em Fast Transforms Algorithms, Analyses, Applications}.
\newblock Academic Press, 1982.

\bibitem{ref4}
J.~H. McClellan.
\newblock Comments on eigenvalue and eigenvector decomposition of the discrete
  {F}ourier transform.
\newblock {\em {IEEE} Trans. Audio Electroacoust.}, 21:65, Feb. 1973.

\bibitem{ref3}
J.~H. McClellan and T.~W. Parks.
\newblock Eigenvalue and eigenvector decomposition of the discrete {F}ourier
  transform.
\newblock {\em {IEEE} Trans. Audio Electroacoust.}, 20:66--74, Jan. 1972.

\bibitem{ozaydin2006orthogonal}
M.~Ozaydin, S.~Nemati, M.~Yeary, and V.~DeBrunner.
\newblock Orthogonal projections and discrete fractional {F}ourier transforms.
\newblock In {\em Proceedings of the 12th Digital Signal Processing Workshop
  and 4th Signal Processing Education Workshop}, pages 429--433, Teton National
  Park, WY, Sept. 2006.

\bibitem{ref2}
S.-C. Pei and J.-J. Ding.
\newblock Eigenfunctions of linear canonical transform.
\newblock {\em {IEEE} Trans. Signal Process.}, 50:11--26, Jan. 2002.

\bibitem{pei2004eigenvectors}
S.-C. Pei and J.-J. Ding.
\newblock Generalized eigenvectors and fractionalization of offset {DFTs} and
  {DCTs}.
\newblock {\em {IEEE} Trans. Signal Process.}, 52(7):2032--2046, July 2004.

\bibitem{pei2008generalized}
S.-C. Pei, J.-J. Ding, W.-L. Hsue, and K.-W. Chang.
\newblock Generalized commuting matrices and their eigenvectors for {DFTs},
  offset {DFTs}, and other periodic operations.
\newblock {\em {IEEE} Trans. Signal Process.}, 56(8):3891--3904, Aug. 2008.

\bibitem{pei2009arbitrary}
S.-C. Pei, W.-L. Hsue, and J.-J. Ding.
\newblock {DFT}-commuting matrix with arbitrary or infinite order second
  derivative approximation.
\newblock {\em {IEEE} Trans. Signal Process.}, 57(1):390--394, Jan. 2009.

\bibitem{ref9}
C.~C. Tseng.
\newblock Eigenvalues and eigenvectors of generalized {DFT}, generalized {DHT},
  {DCT}-{IV} and {DST}-{IV} matrices.
\newblock {\em {IEEE} Trans. Signal Process.}, 50:866--877, Apr. 2002.

\bibitem{yoshida1993blur}
Y.~Yoshida, K.~Horiike, and K.~Fujita.
\newblock Parameter estimation of uniform image blur using {DCT}.
\newblock {\em IEICE Transactions on Fundamentals of Electronics,
  Communications and Computer Sciences}, E76-A(7):1154--1157, July 1993.

\end{thebibliography}
}

\end{document}